\documentclass[12pt,preprint]{aastex}
\usepackage{t1enc}
\usepackage{graphicx}
\usepackage{epstopdf}
\usepackage{natbib}
\usepackage{lscape}
\newcommand{\kms}{km~s$^{-1}$}

\newcommand{\msun}{$M_{\odot}$}
\newcommand{\lsun}{$L_{\odot}$}

\shorttitle{Study of the dense core ahead of HH 80N}
\shortauthors{Masqu\'e et al.}

\begin{document}

\title{Centimeter continuum observations of the northern head of the HH 80/81/80N jet: revising the actual dimensions of a parsec scale jet}

\author{Josep M. Masqu\'e\altaffilmark{1}, Josep M.
Girart\altaffilmark{2}, Robert
Estalella\altaffilmark{1}, Luis F. Rodr\'iguez\altaffilmark{3} \and Maria T. Beltr\'an\altaffilmark{4}}

\altaffiltext{1}{Departament d'Astronomia i Meteorologia, Universitat de Barcelona, 
Mart\'i i Franqu\`es 1, 08028 Barcelona, Catalunya, Spain}

\altaffiltext{2}{Institut de Ci\`encies de l'Espai, (CSIC-IEEC), 
Campus UAB, Facultat de Ci\`encies, Torre C5 - parell 2, 
08193 Bellaterra, Catalunya, Spain}

\altaffiltext{3}{Centro de Radioastronom\'ia y Astrof\'isica, Universidad Nacional Aut\'onoma de M\'exico, Apartado Postal 3-72, 58090 Morelia, Michoac\'an, M\'exico}

\altaffiltext{4}{INAF - Osservatorio Astrofisico di Arcetri, Largo E. Fermi 5, 50125 Firenze, Italy}

\begin{abstract}

We present 6 and 20~cm JVLA/VLA observations of the northern head of the HH 80/81/80N jet, one of the largest collimated jet systems known so far, aimed to look for knots further away than HH 80N, the northern head of the jet. Aligned with the jet and 10\arcmin\ northeast of HH 80N, we found a radio source not reported before, with a negative spectral index similar to that HH 80, HH 81 and HH 80N. The fit of a precessing jet model to the knots of the HH 80/81/80N jet, including the new source, shows that 
the position of this source is close to the jet path resulting from the modeling. If the new source belongs to the HH 80/81/80N jet, its derived size and dynamical age are 18.4~pc and $>9 \times10^3$~yr, respectively. 
If the jet is symmetric, its southern lobe would expand beyond the cloud edge resulting in an asymmetric appearance of the jet. Based on the updated dynamical age, we speculate on the possibility that the HH 80/81/80N jet triggered the star formation observed in a dense core found ahead of HH 80N, which shows signposts of interaction with the jet. These results indicate that pc scale radio jets can play a role on the stability of dense clumps and the regulation of star formation in the molecular cloud.

\end{abstract}

\section{Introduction}

Expanding Herbig-Haro (HH) flows arising from protostellar objects are among the most
spectacular observed phenomena during the star formation process  \citep{herbig1981}. HH flows, which can extend over parsec scale distances \citep{poetzel1989}, are signposts of interaction of the protostellar activity with the ambient cloud medium. At scales of a fraction of parsec from the powering protostar, they appear collimated showing  a jet-like morphology   \citep{mundt1987}. However, at parsec scales the collimation is less evident because first, the possible precession can enhance an S-shaped morphology for the flow and, second, the common presence of multiple outflow sources in star forming clouds may cause contamination by neighboring flows  \citep{bally1994,reipurth1997}. As a consequence, the measurement of the actual length of HH flows can be sometimes controversial, since it is not easy to associate distant and
tentative HH  objects to a specific HH flow without further evidence (e.g. through
proper motions or physical properties of the object).

A peculiar case, the HH 80/81/80N jet, is found in the region GGD 27 \citep{gyulbudaghian1978}, located in  Sagittarius at a distance of 1.7~kpc \citep{rodriguez1980}. The jet is powered  by a
young, high-luminosity protostellar object, IRAS 18162$-$2048 with a total luminosity of $1.7 \times 10^4$~\lsun,  derived from the IRAS survey. Recently,  \citet{fernandezlopez2011a,fernandezlopez2011b}  detect compact millimeter  emission towards this source, which is interpreted by these authors as  arising from a massive ($\sim4$~\msun), and compact ($r$~$\la$~300~AU) disk. VLA observations carried out towards IRAS 18162$-$2048 by \citet{rodriguez1989} reveal that
this source is elongated, pointing to the south toward HH 80 and 81, the brightest Herbig-Haro objects known. HH 80N is located 3~pc north of IRAS~18162$-$2048 and is the northern counterpart of HH 80 and 81. HH 80N has only been detected at radio wavelengths due to the possible high extinction of this region \citep{marti1993}. These authors have also found several radio knots aligned with the central source and the HH 
objects, which indicates that these condensations probably trace a highly collimated jet. Evidence of wiggling of the flow axis suggests that the driving source is precessing, and preserving the collimation up to scales of $\sim$~5~pc, 
being one of the largest collimated jet systems known so far \citep{marti1993}. The high outflow velocities of 500~km~s$^{-1}$, derived from proper motions measurements \citep{marti1995,marti1998}, suggest that the emission arising from the radio jet is due to very strong shocks \citep{heathcote1998}. Synchrotron  radiation, indicating the presence of relativistic electrons, has been recently  found in this jet \citep{carrascogonzalez2010}.  

In this letter we report the results of recent cm continuum observations carried out with the Jansky Very Large Array (JVLA)/Very Large Array (VLA) toward the northern lobe of the HH 80/81/80N jet aimed at looking for knots further away than HH 80N. 
Up to date,  HH 80N is assumed to be the head of the HH 80/81/80N jet. Determining the true size and age of the jet is crucial, as it 
can be an important supplier of energy and momentum to the cloud. Therefore, the HH 80/81/80N jet is a good target to investigate the effects of pc scale jets on star formation in molecular clouds. 

\section{Observations and results} 

JVLA/VLA continuum observations of the northern part of the HH 80/81/80N radio-jet
were carried out on June 2009 in the C configuration (6 and 20~cm bands) and October 2009 in the D configuration (6~cm band). The phase center was always set at the position of the HH 80N object, $\alpha(J2000) = 18^\mathrm{h} 19^\mathrm{m} 19\fs74$
and $\delta(J2000) = -20 \degr 41\arcmin 34\farcs9$, except for the D configuration observations, for which we performed a two
point mosaic with one field centered on the HH 80N position and the other offset (81\arcsec, 328\arcsec). 
A bandwidth of 100~MHz (for the two polarizations) was employed. The flux calibrator was always 3C286 and the phase calibrators were J1820$-$254 at 6~cm and J1833$-$210 at 20~cm. The data were
edited and calibrated using the AIPS package of NRAO.

\begin{landscape}
 \begin{figure}[htbp]
\begin{center}
\resizebox{1.2\textwidth}{!}{\includegraphics[angle=270]{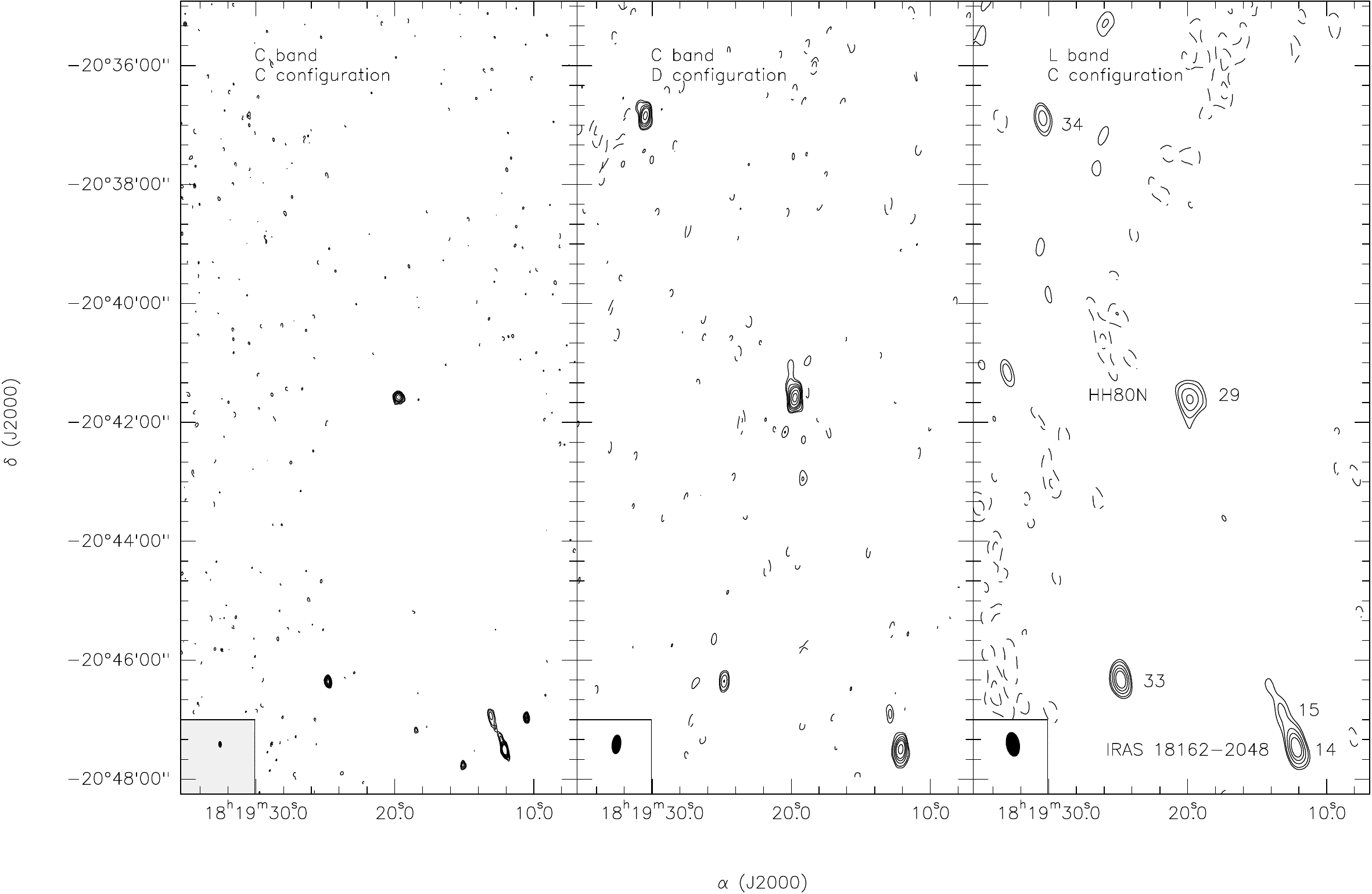}}
\caption{Northern lobe of the HH 80/81/80N jet mapped with centimeter continuum emission.
The left panel shows the 6~cm continuum map obtained with C configuration. The center panel shows the 6~cm continuum map obtained with D configuration. The right panel shows the
20~cm continuum map obtained with C configuration. Contour levels are -3, 3, 4, 6, 8, 10, 14, 18, and
25 times 40, 65 and 250~$\mu$Jy, the rms noise of the left, middle and right maps, respectively. The beam
is shown in the bottom left corner of each panel:
6.6\arcsec $\times$ 3.5\arcsec\ and P.A. of  -2.7\arcdeg\ (\emph{left panel}),  
19.9\arcsec $\times$ 9.7\arcsec\ and P.A. of 10.0\arcdeg\ (\emph{center panel}), and
25.1\arcsec $\times$ 14.0\arcsec\ and P.A. of -2.9\arcdeg\  (\emph{right panel}).
The numbers on the right panel label the sources found in the field following the \citet{marti1993} nomenclature.
\label{VLA_cm}}
\end{center}
\end{figure}
\end{landscape}

The maps were highly contaminated by the emission of a strong radio source, J1819$-$2036, located at $\alpha(J2000) = 18^\mathrm{h} 19^\mathrm{m} 36\fs9$ and $\delta(J2000) = -20 \degr 36\arcmin 31\farcs0$. 
 This source was first reported by \citet{furst1990} from a radio continuum survey at 11~cm obtained with the
Effelsberg~100m antenna (their source num. 359). The flux density measured in our 6 and 20~cm maps is $113.0 \pm 0.3$ and $202.5 \pm 0.2$~mJy, respectively. In order to avoid the strong sidelobes generated by this source,  its clean components were subtracted from the visibility dataset of each block. 
The resulting final maps of the GGD~27 region are shown in Figure~\ref{VLA_cm}. 

Figure~\ref{VLA_cm} shows that IRAS~18162$-$2048 is elongated, pointing to HH 80N. A few marginal knots
appear in the map connecting both objects following the shape of a jet. This is best seen in the 6~cm map obtained with the D configuration. The 6~cm map in the C configuration proves that the jet is very well collimated. At 20~cm, contamination at short
baselines, probably due to Galactic background emission, prevented the proper imaging of the extended emission.

Table~\ref{VLA_fluxes} gives the measured position, flux at 6 and 20~cm, and the spectral index of the sources
belonging to the HH 80/81/80N radio-jet. Our flux values are systematically slightly lower than those reported in Table~2 of
\citet{marti1993}, possibly due to calibration uncertainties. The spectral indices are consistent with those derived by
\citet{marti1993} except for HH 80N, for which we found a spectral index slightly more negative.
All spectral indices are well below the minimum value required for having free-free emission (-0.1, \citealt{rodriguez1993}), which suggests that the contribution of synchrotron emission is significant \citep{carrascogonzalez2010}.  Source 33 has very negative spectral index ($\sim-0.8$). This source is not aligned with the jet path and is likely extragalactic.

Source 34, detected at the 6 and 20~cm bands with no associated counterparts at any other wavelength, appears to be located  about 6\arcmin\ north of HH 80N
($\sim10$\arcmin\ from IRAS 18162-2048), roughly in the direction of the HH 80/81/80N radio jet. 
This source, not reported before, has a negative spectral index of $-0.46$, similar to the values found
for HH 80, HH 81 and HH 80N, and characteristic of the jet. The D configuration map at 6~cm (Fig.~1) shows that  this source is elongated in the north-south direction aligned roughly with the direction of the jet. What is the a priori probability that Source 34 is a background source
unrelated to the star forming region? We will assume that we would have
considered the source associated with the outflow if it appeared anywhere
inside a rectangular box with dimensions of $0\rlap.{'}5 \times 10{'}$
with its major dimension along the outflow axis. The solid angle of
this rectangle is 5 square arcmin. Following \citet{fomalont1991} the a priori probability of finding a 1~mJy background source at 6 cm inside
such a solid angle is only $\sim$0.03. We then consider unlikely that
Source 34 is not associated with the star forming region.

\citet{marti1993} found that the different radio knots of the HH 80/81/80N radio jet are not completely
aligned but rather follow locations compatible with a precessing jet.  In order to test if Source 34 is part of the HH 80/81/80N jet, we extrapolated the expected path of the jet modeled by \citet{marti1993} assuming that the jet is precessing with the jet axis in the plane of sky. Using the coordinates $(x,y)$, with $y$ perpendicular to the jet axis and $x$ along the jet axis, the $y$ position of a knot is given by the sinusoidal expression.

\begin{equation} 
y = x \tan(\beta)\cos(\varphi_p-2\pi\frac{|x|}{\lambda_p}) 
\end{equation}

where $\beta$ is the half-opening angle of the jet, $\varphi_p$ is the initial precession phase angle for $x=0$, and $\lambda_p$ is the wavelength of the wiggles in the plane of sky.

\begin{figure}[htbp]
\begin{center}
\resizebox{0.53\textwidth}{!}{\includegraphics{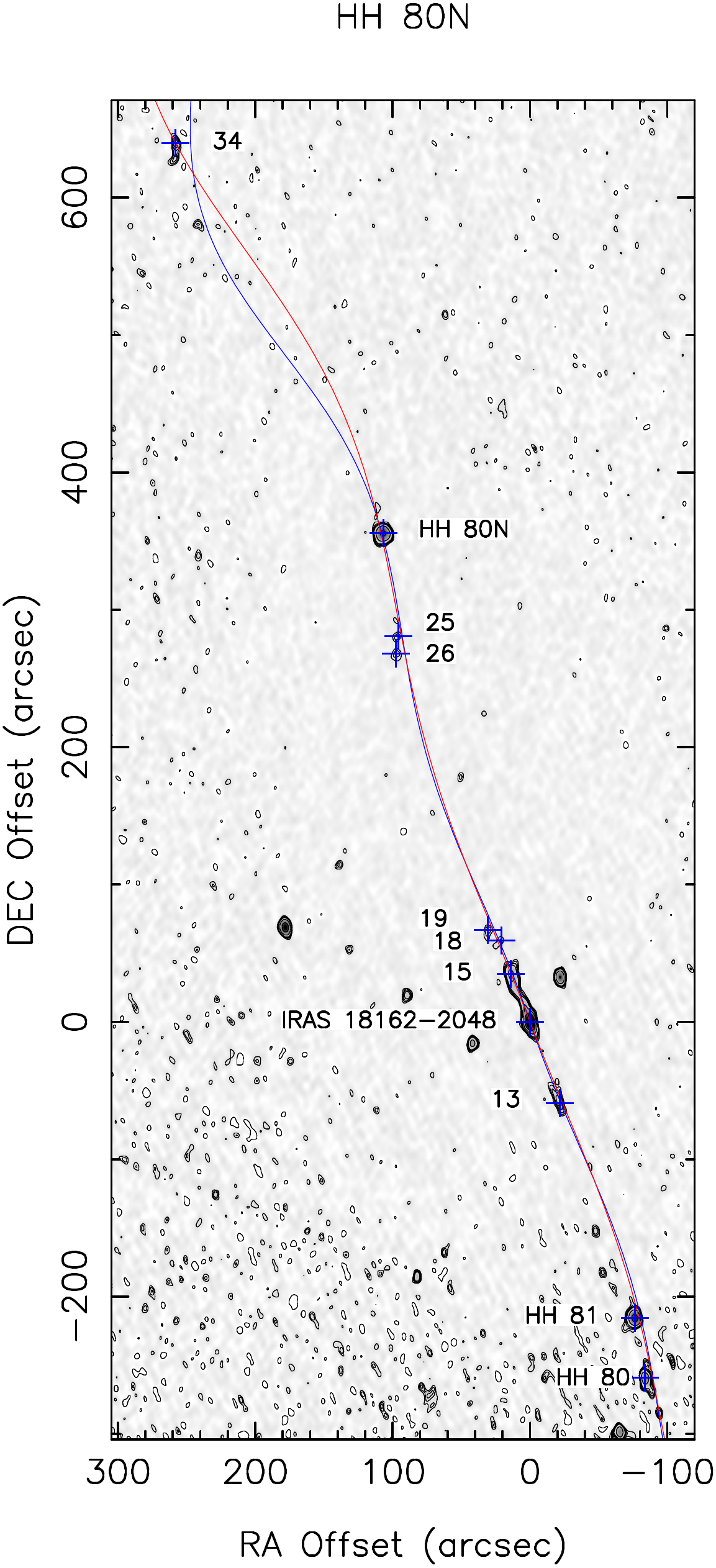}}
\caption{Fit of a precessing jet to the knots of the HH 80/81/80N complex (blue crosses) plotted over the 6 cm map obtained with the C~configuration (including \citet{marti1993} data) convolved with a Gaussian of 15\arcsec\ of FHWM. The numbers follow the \citet{marti1993} nomenclature and indicate the knots used in the fit of a precessing jet (see text). The blue solid line shows the precessing jet path obtained using the parameters of model A (without including Source 34) while the red solid line shows the resulting jet path using the parameters of model B (including Source 34). The reference position of the map is located at IRAS 18162$-$2048.
\label{jiggling}}
\end{center}
\end{figure}

Table~\ref{jiggling_results} gives the results of the fit of the expression (1) to the knots of the HH 80/81/80N jet, without including Source 34 (i.e. aimed at reproducing the results of \citet{marti1993}, model A) and including this source (model B). 
In this analysis we also employed \citet{marti1993} data obtained from observations in the C configuration at 6~cm, which have a field of view covering the jet from HH 80N to HH 80/81, in order to include Source 13,  HH 81 and HH 80. The resulting sinusoidal path for both models are shown in Figure~\ref{jiggling}.  
This figure shows the map resulting from the observations reported in \citet{marti1993} combined with the new data reported in this letter obtained in the C configuration at 6~cm. The parameters of Model A have values similar to those obtained in \citet{marti1993} as expected (note that the precession angle in  \citet{marti1993} is the full opening angle, i.e. $2\beta$). However, the sinusoidal path resulting from this fit misses Source 34 by about 20\arcsec\ (see blue line of Fig.~\ref{jiggling}). On the other hand, model B yields a jet path close to Source 34 (see red line of Fig.~\ref{jiggling}). The parameters of both models have similar values except for $\varphi_p$ and $\lambda_p$. As seen in  Fig.~\ref{jiggling}, in the model B, the jet has longer wiggles in the plane of the sky with respect to model A.

\section{Discussion}

 Given the symmetry of the HH 80/81/80N radio knots, it is plausible that Source~34 has a counterpart in the southern lobe. We inspected the 20~cm map but did not detect any emission above 1.6 mJy at the expected counterpart position. Note that this is not a stringent upper limit due to the contamination at low baselines caused probably by galactic extended emission. We also checked surveys observed at wavelengths suited to detect HH shock emission 
(e.g. H$\alpha$ SuperCosmos) but found no emission around the expected counterpart position. Also, this position falls outside the field of view of the 6 cm band observations. 
Nevertheless, observations of HH complexes show that they are not necessarily strictly symmetric (HH~111: \citealt{rodriguez1994}; HH~34: \citealt{stapelfeldt1991, anglada1995}). The asymmetry of the HH 80/81/80N jet could be caused by  the fact that the southern lobe of the jet moves beyond the edge of the cloud, changing the flow appearance further away than HH 80 because it enters in a low pressure medium. The projected distance from HH 80N to the southern bow shock beyond HH 80 at a distance of 1.7~kpc is 7.5~pc  \citep{heathcote1998}. If Source 34 belongs to the HH 80/81/80N jet then its total extent to the plane of sky is 10.3~pc. Correcting from inclination effects using $56\arcdeg$ for the value of the angle of the jet axis to the plane of the sky  \citep{heathcote1998}, we derive 18.4~pc of total extent of the HH system. This updated size is 5.0~pc larger than the size previously stablished by \citet{heathcote1998}.

The extraordinary collimation of the HH 80/81/80N jet extending from the new reported Source 34 to HH 80 and 81 is in agreement with the idea that the ejection mechanism of the jet operates similarly as in the low  luminosity protostars. However, the results presented here show that the collimation and symmetry of pc scale jets tend to break at some point because 
the chances of finding a heterogeneous medium are higher as the jet propagates far away in the 
cloud. IRAS 18162$-$2048 also reveals similarities with low luminosity protostars. Continuum mm observations 
 towards this source suggest that the emission arises from a massive and compact disk surrounding the driving source of the HH 80/81/80N jet \citep{fernandezlopez2011a,carrascogonzalez2012}. The derived mass ratio between the disk and the central star is $\lesssim$ 0.3 \citep{carrascogonzalez2012},  which seems to be consistent with the typical values found for low mass protostars. These authors also obtained a crude estimate for the mass accretion rate of the disk onto the central powering protostar of $10^{-4}$~\msun~yr$^{-1}$. Adopting a mass for the central object of $7-11$~\msun\ \citep{fernandezlopez2011b}, the accretion phase must have started between $(0.7-1.1) \times 10^{5}$~yr ago.

Assuming that HH 80N is the northern end of the HH 80/81/80N jet and all its condensations 
move ballistically with a constant spatial velocity of 625~\kms\ (obtained using the average tangential velocity of HH 80 and HH 81 and the angle of the jet axis to the plane of sky of \citealt{heathcote1998}), we derive a dynamical age of $\sim5 \times 10^3$~yr for the jet. However, if Source 34 belongs to the HH 80/81/80N jet, the dynamical age increases to $\sim9 \times10^3$~yr. Furthermore, the HH 80/81/80N jet may be decelerating as a consequence of the loss of kinetic energy as the jet passes through the cloud, as indicated by the high mechanical luminosity of HH 80N (a few $10^1$~\lsun,  \citealt{girart1994}). For this case, the dynamical age must be $>9 \times10^3$~yr, consistent with the upper limit for the age of the accretion phase of IRAS 18162$-$2048 derived above.

Ahead of HH 80N there is a dense core extensively studied by our team with signatures of star formation and supersonic infall velocities \citep{girart1994,girart1998,girart2001,masque2009,masque2011}. Given the close location of the core to the most powerful jet known and its peculiar kinematics, it has been suggested that the star formation of this core has been triggered or at least sped up by the HH 80/81/80N jet  \citep{girart2001, masque2011}. The estimated age of the protostar embedded in the HH 80N core  \citep[$2 \times 10^4$~yr,][]{masque2011} falls in the range comprised between the lower limit of the estimated dynamical age of the jet and the upper limit of the age of the main accretion phase of its energy source, $(0.9-11.0) \times 10^4$~yr. 
Moreover, our observations of high density tracers in the HH 80N region (Masqu\'e et al. 2012, in prep.) suggest that there is a dynamical interaction between HH 80/81/80N jet and the HH 80N core: the velocity gradients of the high density gas are consistent with the possibility that part of the core is being swept by the HH 80/81/80N jet in the process of dissipating energy. This could turn some regions in the core gravitationally unstable. This possibility appears also evident in the high angular resolution continuum maps of the HH 80N core \citep{masque2011}, which show that the HH 80N core is fragmenting into several condensations, one of them clearly associated with the embedded object in the core. Therefore, the HH 80/81/80N jet may play a role in the star formation process of the HH 80N core. Conclusive evidence for this possibility would provide an illustrative case of sequential star formation in a cloud occurring at scales of parsecs.

\acknowledgments

JMM, JMG and RE are supported by the Spanish MINECO AYA2011-30228-C03 and the Catalan AGAUR  2009SGR1172 grants.

\begin{table}[ht]
\caption{Parameters of the knots of the HH 80/81/80N jet$^{\mathrm{a}}$\label{VLA_fluxes}}
\begin{tabular}{lccccc}
\hline\hline
                             &   \multicolumn{2}{c}{Peak Position}                                                   & $S_\mathrm{20cm}$ & $S_\mathrm{6cm}$                               \\                                        
Source                &                  $\alpha$~(J2000)                                               &       $\delta$~(J2000)          & (mJy)     & (mJy) & Spectral Index            \\ 
\hline 
7$^{\mathrm{b,c,d}}$        &     $18^\mathrm{h}19^\mathrm{m}06\fs122$       &       $-20\degr51'49\farcs486$   & $16.83 \pm 0.07$$^{\mathrm{e}}$ & $0.77\pm0.06$&    $\geq -2.47$  \\
8$^{\mathrm{b,c,f}}$        &     $18^\mathrm{h}19^\mathrm{m}06\fs652$       &       $-20\degr51'05\farcs822$   &            ''                    & $1.54\pm0.07$&    $\geq -1.91$ \\
13$^{\mathrm{c}}$        &     $18^\mathrm{h}19^\mathrm{m}10\fs554$       &       $-20\degr48'29\farcs239$   & $0.69 \pm $ 0.06 & $0.52\pm0.10$&    $-0.22 \pm 0.17$  \\
14$^{\mathrm{g}}$         &   $18^\mathrm{h}19^\mathrm{m}12\fs106$         &     $-20\degr47'30\farcs834$    &$3.77\pm0.10$ & $4.29\pm0.06$&  $0.11\pm0.01$    \\
15$^{\mathrm{b,c}}$        &     $18^\mathrm{h}19^\mathrm{m}13\fs086$       &       $-20\degr46'55\farcs713$   & $1.09 \pm 0.07$ & $1.06\pm0.10$&    $-0.02 \pm 0.09$  \\
29$^{\mathrm{h}}$        &     $18^\mathrm{h}19^\mathrm{m}19\fs780$       &       $-20\degr41'33\farcs343$    &$3.72\pm0.20$ & $2.05\pm0.15$&  $-0.49\pm0.04$    \\
34$^{\mathrm{i}}$        &     $18^\mathrm{h}19^\mathrm{m}30\fs489$       &       $-20\degr36'51\farcs315$    &$1.82\pm0.05$ & $1.03\pm0.03$&   $-0.46\pm0.02$   \\
\hline 
\end{tabular}
\vspace{0.5cm}\\
\textbf{Notes.} \\
$^{\mathrm{a}}${Derived from a Gaussian fit}\\
$^{\mathrm{b}}${Sources with position outside the field of view of Fig.~\ref{VLA_cm}}\\
$^{\mathrm{c}}${6~cm fluxes were obtained using \citet{marti1993} data at C configuration}\\
$^{\mathrm{d}}${HH 80}\\
$^{\mathrm{e}}${Sources 7 and 8 appear blended in the 20~cm map}\\
$^{\mathrm{f}}${HH 81}\\
$^{\mathrm{g}}${IRAS $18162-2048$}\\
$^{\mathrm{h}}${HH 80N}\\
$^{\mathrm{i}}${New reported source}\\
\end{table}

\begin{table}[bh]
\vspace{-0.5cm}
\centering
\caption{Fit results of a precessing jet to the knots of the HH 80/81/80N radio jet\label{jiggling_results}}
\begin{tabular}{cccccc}
\hline\hline
                   &       Rms fit residual           & $\beta$                       &     $\lambda_p$         &                                       $\psi$                         &        $\varphi_p$        \\                                        
  Model      &          (arcsec)                      &       (deg)                      &  (arcsec)                   &                                   (deg)                               &                     (rad)         \\
\hline
     A           &            3.5                             &      $2.7 \pm 0.4$        &  $470 \pm 30$          &                                     $19.5 \pm 0.3$            &     $4.9 \pm 0.4$         \\
     B           &            3.2                             &      $2.7 \pm 0.5$        &  $620 \pm 40$          &                                    $19.4 \pm 0.4$             &     $4.2 \pm 0.3$         \\
\hline
\end{tabular}
\end{table}

 \end{document}